\documentclass[12pt]{article}
\usepackage{multicol}
\usepackage{graphicx}
\usepackage{threeparttable}
\usepackage[round]{natbib}
\usepackage{amsfonts}
\usepackage{amsmath}
\usepackage{epstopdf}

\newcommand\lesssim{\mathrel{\hbox{\rlap{\hbox{\lower4pt\hbox{$\sim$}}}\hbox{$<$}}}}
\title{{\Huge NANOGrav}\\ {\Large The North American Nanohertz Observatory for Gravitational Waves}\\ {\large An Astro2010 decadal survey activity submission}}
\begin{document}
\begin{titlepage}
\thispagestyle{empty}
\begin{center}
{\Large\bfseries%
The North American Nanohertz Observatory for Gravitational Waves\\
}
\end{center}

\begin{center}
{\large 
An Astro2010 decadal survey activity submission}
\end{center}


\begin{center}
\includegraphics[width=1.0\textwidth]{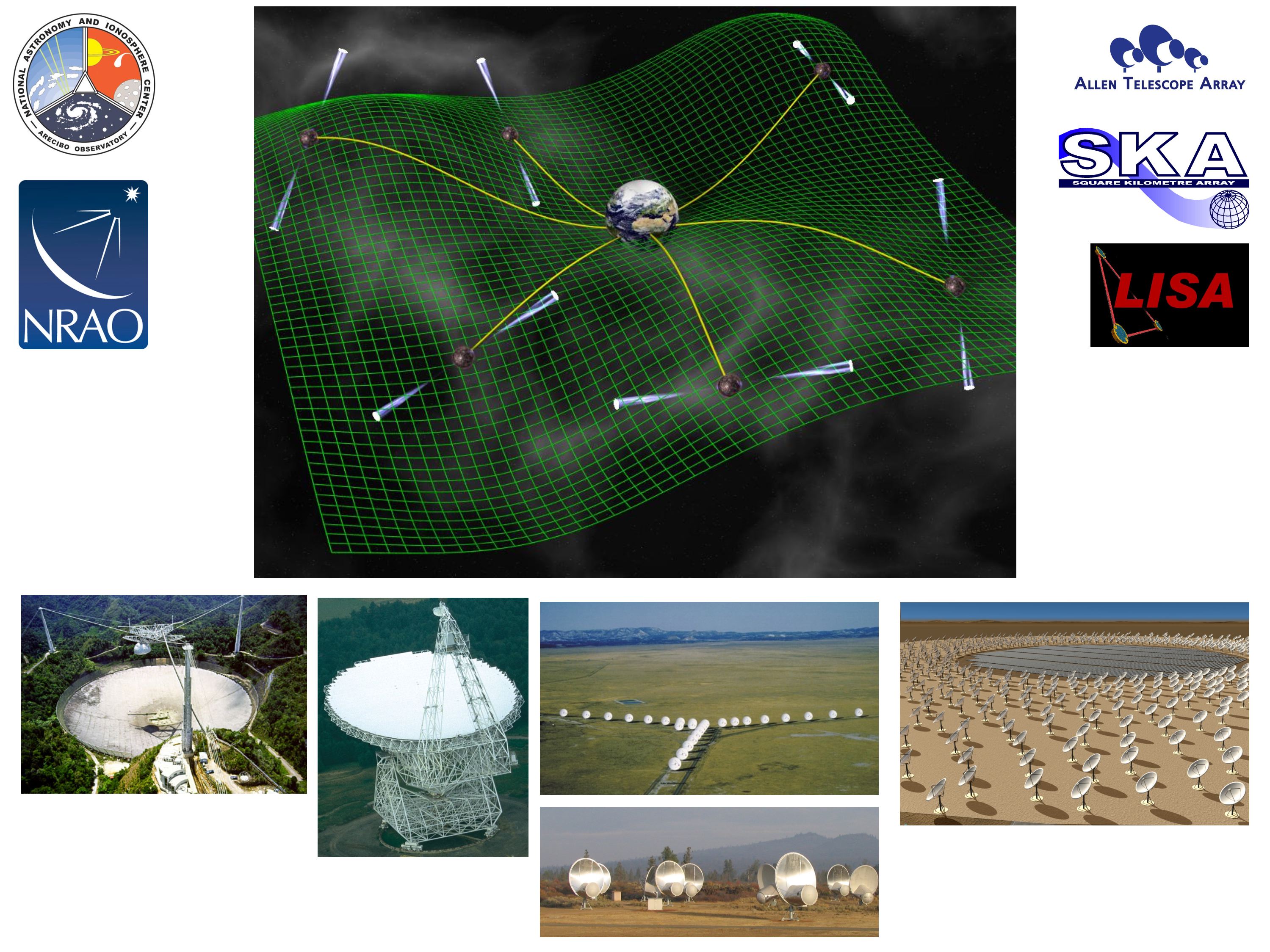}
\end{center}
{\bf Principal Authors:}
F.~Jenet (Center for Gravitational Wave Astronomy @ UT Brownsville, 626-524-0518, \texttt{merlyn@phys.utb.edu});
L.~S.~Finn (Penn State, 814-863-9598, \texttt{lsfinn@psu.edu});  
J.~Lazio (NRL, 202-404-6329, \texttt{Joseph.Lazio@nrl.navy.mil});
A.~Lommen (Franklin and Marshall College, 717-291-3810, \texttt{alommen@fandm.edu});
M.~McLaughlin (West Virginia University, 304-293-3422, \texttt{maura.mclaughlin@mail.wvu.edu});
I.~Stairs (UBC, 604-822-6796, \texttt{stairs@astro.ubc.ca});
D.~Stinebring (Oberlin, 440-775-8331, \texttt{dan.stinebring@oberlin.edu});
J.~Verbiest (WVU, 304-293-3422x1462, \texttt{Joris.Verbiest@mail.wvu.edu});\\
{\bf NANOGrav Members and Contributors:}
A.~Archibald (McGill);
Z.~Arzoumanian (CRESST/USRA/NASA-GSFC);
D.~Backer (UC Berkeley);
J.~Cordes (Cornell);
P.~Demorest (NRAO);
R.~Ferdman (CNRS, France);
P.~Freire (NAIC);
M.~Gonzalez (UBC);
V.~Kaspi (McGill);
V.~Kondratiev (WVU);
D.~Lorimer (WVU);
R.~Lynch (Virginia);
D.~Nice (Bryn Mawr);
S.~Ransom (NRAO);
R.~Shannon (Cornell);
X.~Siemens (UW Milwaukee);

\end{titlepage}
\tableofcontents

\newpage
\section{Summary}

The North American Nanohertz Observatory for Gravitational Waves (NANOGrav) is a consortium of astronomers whose goal is the creation of a galactic scale gravitational wave observatory sensitive to gravitational waves in the nHz -- $\mu$Hz band. It is just one component of an international collaboration involving similar organizations of European and Australian astronomers who share the same goal. 

Gravitational waves, a prediction of Einstein's general theory of relativity, are a phenomenon of dynamical space-time generated by the bulk motion of matter (e.g., the rapid periapsis passage of star on a low angular momentum orbit about a supermassive black hole), primordial or quantum fluctuations arising from early universe phenomena (e.g., cosmic strings, superstrings, or inflation), and the dynamics of space-time itself (e.g., a coalescing black hole binary and its coalescence to form a single black hole). 
The are detectable by the small disturbance they cause in the light travel time between some light source and an observer. NANOGrav exploits radio pulsars as both the light (radio) source and the clock against which the light travel time is measured. In an array of radio pulsars gravitational waves manifest themselves as correlated disturbances in the pulse arrival times. The timing precision of today's best measured pulsars is less than 100~ns. With improved instrumentation and signal-to-noise it is widely believed that the next decade could see a pulsar timing network of 100 pulsars each with better than 100~ns timing precision. Such a \emph{pulsar timing array} (PTA), observed with a regular cadence of days to weeks, would be capable of observing supermassive black hole binaries following galactic mergers, relic radiation from early universe phenomena such as cosmic strings, cosmic superstrings, or inflation, and more generally providing a vantage on the universe whose revolutionary potential has not been seen in the 400 years since Galileo first turned a telescope to the heavens. 

The parameters that determine the sensitivity of a pulsar timing array to gravitational waves are the number, timing precision, and sky distribution of the array's best pulsars. To achieve its goals, NANOGrav requires improved pulsar timing, with higher cadence and more regular timing of the array pulsars; searches for new, stable millisecond pulsars to enlarge the array size (especially in the northern sky); and improved statistical analyses of timing residuals to increase the robustness of gravitational wave detection at lower signal strength. 

To make its observations, NANOGrav can exploit existing and new radio astronomy telescope infrastructure. The broader bandwidth of modern pulsar back-end instrumentation presents the opportunity to dramatically increase pulsar timing precision. Taking advantage of broad-band observations requires the development of techniques that account for interstellar medium propagation effects, frequency dependent profile changes, polarization calibration errors, and radio frequency interference (RFI) mitigation. With the appropriate computational infrastructure, NANOGrav can take excellent advantage of the proposed phased-array radio telescopes (e.g., the Allen Telescope Array and the Square Kilometer Array) and, in turn,  drive the development of a non-imaging timing mode. Such development will prepare the NANOGrav community to take full advantage of the next generation of radio astronomy hardware.   

\newpage
\section{Key Science Goals}


Gravitation powers and governs the most energetic processes in the
cosmos, from supernovae, to gamma-ray bursts, to quasars, to the
coalescence of supermassive black hole binaries following galaxy
mergers. Gravitational waves associated with these phenomena are the only
direct probe of the central engines that power them. Gravitational
waves are also the most direct diagnostic of the structure of
space-time, and of the processes taking place in the earliest moments
in the life of the Universe. Microhertz or lower frequency gravitational
waves associated with, e.g., coalescing supermassive black holes or
black hole binaries cannot be detected using conventional ground- or
space-based detectors. They can, however, be detected through the
correlated imprint they leave on the timing residuals derived from
observations of an array of pulsars. The combination of expected
source strengths, timing precision and number of low-timing-noise
pulsars suggest strongly that gravitational wave observations with a
pulsar timing array will be possible within the next decade
\citep{jenet:2005:dsg}. The North American Nanohertz Observatory for
Gravitational Waves --- NANOGrav --- is a consortium of
astronomers whose goal is to observe nHz--$\mu$Hz frequency
gravitational waves and use those observations as a tool of
observational astronomy, contributing principally to
\begin{enumerate}
\item Understanding the co-evolution of galaxies and supermassive
  black holes;
\item Searching for signatures of early-universe or exotic physics
  processes (e.g., inflation or cosmic strings);
\item Probing the nature of space-time, including the search for
  quantum gravity corrections to classical gravity; and
\item Discovering sources of gravitational waves previously
  unrecognized.
\end{enumerate}

\subsection{Background}
Pulsars are exquisite clocks. Among all pulsars, the class of
millisecond pulsars are especially precise timekeepers. Owing their
stability to their large moment of inertia and ``seismic''
stability, the rms timing residuals --- the delays between the expected
and actual arrival time of pulses --- for several millisecond pulsars
are below 100~ns. The timing residuals have steadily improved with
improved instrumentation and signal-to-noise, strongly suggesting that
we have not yet reached the intrinsic pulsar timing noise
limits. 

A pulsar's regular tempo makes it an invaluable tool for exploring
relativistic phenomena where precise and accurate measurement of
duration or interval is critical.  The \citet{hulse:1975:dop}
discovery of the pulsar PSR~B1913$+$16 in a binary system enabled the
precision measurement of multiple relativistic corrections to
Newtonian gravity, culminating with a measurement of the orbital period decay rate with precision $10^{-15}$~s/s \citep{weisberg:2005:rbp}. The agreement between 
this $\dot{P}_{\mathrm{b}}$ and general relativity's prediction for the orbital period
evolution owing to gravitational wave emission --- currently 0.2\% --- was responsible for the
first and, currently, best observational evidence for gravitational
wave emission. The agreement between measured binary period decay and the
predictions of general relativity provided by PSR~B1913$+$16 and
other, subsequently discovered binary systems is at this writing the
\emph{only} test of general relativity theory in its dynamical sector.

In a binary pulsar system gravitational wave emission is inferred from
its dissipative effect on the binary's orbital period. The pulsar is,
in this case, the precision clock that enables us to characterize the
evolving orbit. Gravitational waves also affect space-time: in particular,
they disturb the time it takes a pulse to propagate between a pulsar
and a radio telescope observatory. The passing wave's effect on this
propagation time depends on the wave propagation direction and
polarization relative to the pulsar-observatory line-of-sight.
Correlations among the timing residuals measured for an array of
pulsars along different lines-of-sight can thus reveal the presence
of gravitational waves, their propagation direction and their
polarization.

The pulsar timing residual associated with gravitational waves
crossing the observatory-pulsar line-of-sight is equal to geometrical
factors times the time-integrated gravitational wave strain, commonly
denoted $h_{ij}$, along the electromagnetic wave-front as it propagates
from the pulsar to the observatory. \emph{The ability to use the pulsar timing array to 
measure the equivalent of nanosecond or less timing residuals induced by gravitational waves, corresponding to a strain sensitivity to broadband bursts of order $h\sim6\times10^{-15}(f_{\mathrm{gw}}/10^{-6}~\mathrm{Hz})$, is well within the capability of next-decade pulsar timing array measurements.} The subsections below describe how NANOGrav will exploit this sensitivity to achieve its principal science goals.

\subsection{Understand the co-evolution of galaxies and supermassive black 
holes}\label{sec:coevol}




Massive black holes appear to be ubiquitous inhabitants of all
galactic nuclei with spheroids \citep{richstone:1998:sbh}. The
properties of these black holes also appear to be strongly correlated
with the properties of their hosts
\citep{magorrian:1998:dom,gebhardt:2000:bhm,ferrarese:2000:frb}. Together
these observations support a co-evolution scenario for massive black
holes and galaxies that would extend to other phenomena (e.g., energy 
radiated by a quasar over its lifetime) previously regarded as independent
\citep{hopkins:2008:cff}. In this scenario, the correlations between
nuclear black holes and their hosts are driven by the processes that
follow from galaxy mergers. Following a merger, dynamical friction
leads the massive black holes to sink to the center of the merger
remnant. At the center they form a bound binary that loses its orbital
energy through interactions with stars and gas. Eventually the stars
and gas are exhausted and the coalescence is driven by gravitational
wave emission \citep{begelman:1980:mbh}.

The timing residual for a massive black hole binary is readily
estimated. For a binary with total mass $\mathrm{M}/(1+z)$ at luminosity
distance $\mathrm{d_L}$ radiating gravitational waves with observed frequency
$f$ the induced timing residual magnitude is, up to geometric factors,
\begin{equation}
  \Delta\tau \sim 10\,\mathrm{ns}\,
  \left(\frac{1\,\mathrm{Gpc}}{\mathrm{d_L}}\right)
  \left(\frac{\mathrm{M}}{10^9\,\mathrm{M}_\odot}\right)^{5/3}
  \left(\frac{10^{-7}\,\mathrm{Hz}}{f}\right)^{1/3}
\end{equation}
Correlated timing residuals of this magnitude are well within reach of
pulsar timing array measurements in the next decade. 

Observations with an array of 20 pulsars, each with 100~ns rms timing
residual noise, are expected to observe a confusion-limited
gravitational wave ``background'' signal in the nHz--$\mu$Hz window
arising from binaries with masses $>10^8\,\mathrm{M}_{\odot}$
\citep{jaffe:2003:gwp,jenet:2005:dsg,sesana:2008:sgb}. On top of this
confusion-limited signal should be a handful of individually resolvable
binaries, of greater mass, at distances $z\lesssim2$
\citep{sesana:2008:gwf}. The level and spectral shape of the background
signal reflects the abundance of massive binaries and their
eccentricity, which are determined by the dynamical processes acting
at sub--parsec scales.
The background spectral shape is also strongly constrained by the still
disputed shape of the high-mass end
of the black hole mass function (e.g., \cite{lauer:2007:s}). Observations of the gravitational
wave background in this band will thus provide valuable insights into
the poorly known accretion and feedback processes that govern the
growth of the massive black holes that play a dominant role in shaping
the structure of giant galaxies and galaxy clusters.


The observation of individually resolvable binaries offers even more
exciting prospects. For these systems the binary's location on the
sky, orbital plane inclination, component mass ratio and total mass to
luminosity distance ratio can all be determined. The mass ratio will
be especially valuable for shaping our understanding of galaxy mergers
and galaxy formation processes. If an electromagnetic counterpart can
also be identified the binary system's component masses will be
determined (from the mass/distance ratio) and a unique laboratory for
studying accretion physics and the interplay between black holes and
their host galaxies uncovered.

\subsection{Search for the signatures of early-universe or exotic physics processes}\label{sec:earlyU}

Gravitational wave relics of early universe phenomena carry the signature of the physics at work in extreme environments otherwise inaccessible to observation or experiment.  The relic radiation extends over wavelengths ranging from the present-day horizon scale (corresponding to $3\times10^{-18}$~Hz) to a cut-off determined by the Planck scale and the expansion dynamics of the universe (which is expected to be of order gigahertz). Over this broad dynamic range in frequency, cosmic microwave background polarization measurements are sensitive to gravitational waves with wavelengths near the horizon scale, pulsar timing observations to radiation in the nHz--$\mu$Hz band, the proposed Laser Interferometer Space Antenna (LISA) detector to the 0.1--100~mHz band, and the advanced Laser Interferometer Gravitational-Wave Observatory (LIGO) detector to the 10--1000~Hz band. While each of these different means of detecting relic gravitational waves thus explores a very different range of frequencies where the radiation spectrum is dominated by different processes, at present and into the foreseeable future pulsar timing experiments provide the tightest constraints on the parameter space of viable cosmic string and superstring models~\citep{Siemens:2006yp}.

In the nHz--$\mu$Hz band relevant for pulsar timing array gravitational wave detection, the dominant early-universe contributions to the gravitational wave background are cosmic strings, cosmic superstrings, and inflation. 
Cosmic strings are one-dimensional topological defects associated with a symmetry-breaking phase transition \citep{Kibble:1976sj} in the early universe. Originally proposed to provide the seeds of structure formation, by the late 1990's this motivation evaporated when they were found to be inconsistent with the distribution of the large-scale structure and the observation of acoustic peaks in the cosmic microwave background. Nevertheless, cosmic strings, which are a generic phenomenon of supersymmetric grand unified theories \citep{Jeannerot:2003qv}, remain viable as a sub-dominant contributor to the cosmic gravitational wave background \citep{wyman:2005:boc}.
\emph{Superstring}-motivated inflation models can also lead to artifacts that behave like cosmic strings; such artifacts are called cosmic superstrings \citep{Polchinski:2004hb}.  Both cosmic strings and superstrings support oscillations, cusps and kinks, which in turn give rise to a stochastic gravitational wave ``background'' and gravitational wave bursts \citep{damour:2001:gwb,damour:2005:grf}.
Finally, inflation gives rise to a stochastic cosmological gravitational wave background by powering a parametric amplification of gravitational wave zero-point quantum fluctuations \citep{grishchuk:2005:rgw}. 

Remarkably, at $10^{-7}$~Hz current predictions for the contribution of cosmic strings, superstrings or inflation to the stochastic gravitational wave background are all of comparable magnitude ($h_{\mathrm{cs}}\sim10^{-16}$--$10^{-14}$ and $h_{\mathrm{infl}}\sim10^{-17}$--$10^{-15}$) and within the reach of pulsar timing observations in the next decade \citep{jenet:2006:ubo}. The spectral index of each of these contributions differ, offering the prospect that the contributions can be separately identified from the spectrum of the background measured over the nHz--$\mu$Hz bandwidth. (These amplitudes are also comparable with that expected from the confusion limit of supermassive black hole binaries, which has yet a different spectral index allowing it to also be distinguished.)

\subsection{Probe the nature of space-time}\label{sec:tests}

\paragraph{Quantum gravity.} 
Astronomical observations involve the largest imaginable scales of 
distance and time. The leverage provided by the vast scales 
probed by astronomical observations enable tests of physical theories 
that cannot be approached in terrestrial laboratories or by 
conventional experiment. Classic examples include measurements 
bounding possible time-evolution of the fine structure ``constant,'' 
which involve comparison of spectral line multiplet frequencies made 
at present and when the universe was considerably younger 
\citep{darling:2003:mfc,kozlov:2004:svo} and bounds on the mass of the 
photon \citep{adelberger:2007:pbd,chibisov:1976:aul}. In a similar 
fashion, gravitational wave observations of sources at cosmological 
distances will make possible tests of string and other theories that 
attempt to merge quantum mechanics and gravity.

All modern routes leading to a quantum theory of gravity ---e.g.,
perturbative quantum gravitational one-loop exact correction to the
global chiral current in the standard model, string theory, and loop
quantum gravity --- require modification of the classical
Einstein-Hilbert action by the addition of a parity-violating
Chern-Simons term. This quantum correction affects only the
gravitation sector of these theories; it has no effect on
electromagnetism or electromagnetic phenomena. While the correction
may be intrinsically small, its effect on gravitational waves
accumulates secularly as the wave propagates. Observation of
gravitational waves that have propagated over cosmological distances
will be capable of measuring or bounding even a small birefringence,
providing evidence that will help to guide the unification of
gravitation and quantum theories \citep{alexander:2008:gpo}.

\paragraph{General relativity, or beyond?} 
Solar system observations and equivalence principle experiments 
have shown that gravitation is almost certainly described by a metric 
theory \citep{will:2006:cbg}. Einstein's theory of gravity --- general
relativity --- is the simplest such theory. While general relativity
has been extremely successful in describing gravitational physics
there still remain many feasible alternative gravity
theories. Interest in such theories has increased recently due to
discoveries in galactic dynamics and cosmology, i.e., dark matter and
cosmic acceleration (see the review \citep{sanders:2002:mnd}). General
relativity makes an unambiguous prediction for the number of
gravitational wave polarization states (two), which is fewer than
allowed (six) by these viable alternative theories
\citep{eardley:1973:goa,eardley:1973:goa:1}. Thus, gravitational wave
observations that can independently resolve the different polarization
states enable a test that can distinguish between general relativity
and other viable alternative theories of gravity.

Terrestrial and space-based gravitational wave detectors are each
sensitive to a single linear combination of the polarization states
associated with a passing gravitational wave. When data from multiple
independent detectors are joined and analyzed coherently the ability
to resolve the different polarization states is increased. For this
purpose, each pulsar line-of-sight is an independent detector: i.e.,
the timing residuals associated with different pulsars each respond to
a different linear combination of gravitational wave
polarizations. Over the next decade there will likely be no more than
three large independent terrestrial gravitational wave detectors (detectors
at the two LIGO sites, and the French-Italian Virgo detector) and at
most one space-based detector (LISA). There are already about 10
independent pulsar baselines with of order 100~ns timing noise and
good reason to expect there to be more than 100 pulsar baselines with
sub 100~ns timing noise by the end of the next decade. The size of
this ``detector array'', which cannot be matched by ground or
space-based experiment, enables observations capable of distinguishing
radiation modes associated with each polarization state and, thus,
distinguishing between general relativity and other alternative
theories of gravity \citep{lee:2008:tap}.

\subsection{Discover previously unanticipated gravitational wave sources}

Nothing more exemplifies the transformative potential of gravitational
wave astronomy than the possibility of discovering
gravitational waves from unforeseen sources or processes. It is cliche
to say that the history of astronomical observation is a catalog of
surprises as new observational channels create new opportunities for
discovery. Behind the cliche, however, is the truth from which it is
made: the range of our direct experience, from which our knowledge and
understanding of the natural world is formed, is very limited when
compared to the range of scales and environments present in the
universe. For this reason, great leaps in understanding follow from
great leaps in observational capability. This decade the gravitational
wave universe lays claim to the imagination as the great, unexplored
frontier ripe for exploration. NANOGrav, its international partners,
and other gravitational wave experiments, are the means that will
carry us into this new frontier, from which we may enrich our
understanding of the known, and the as yet unknown, cosmos.

\newpage
\section{Technical Overview}
\label{technical}

%

\begin{figure}[tbh]
\begin{center}
\includegraphics[width=0.9\textwidth]{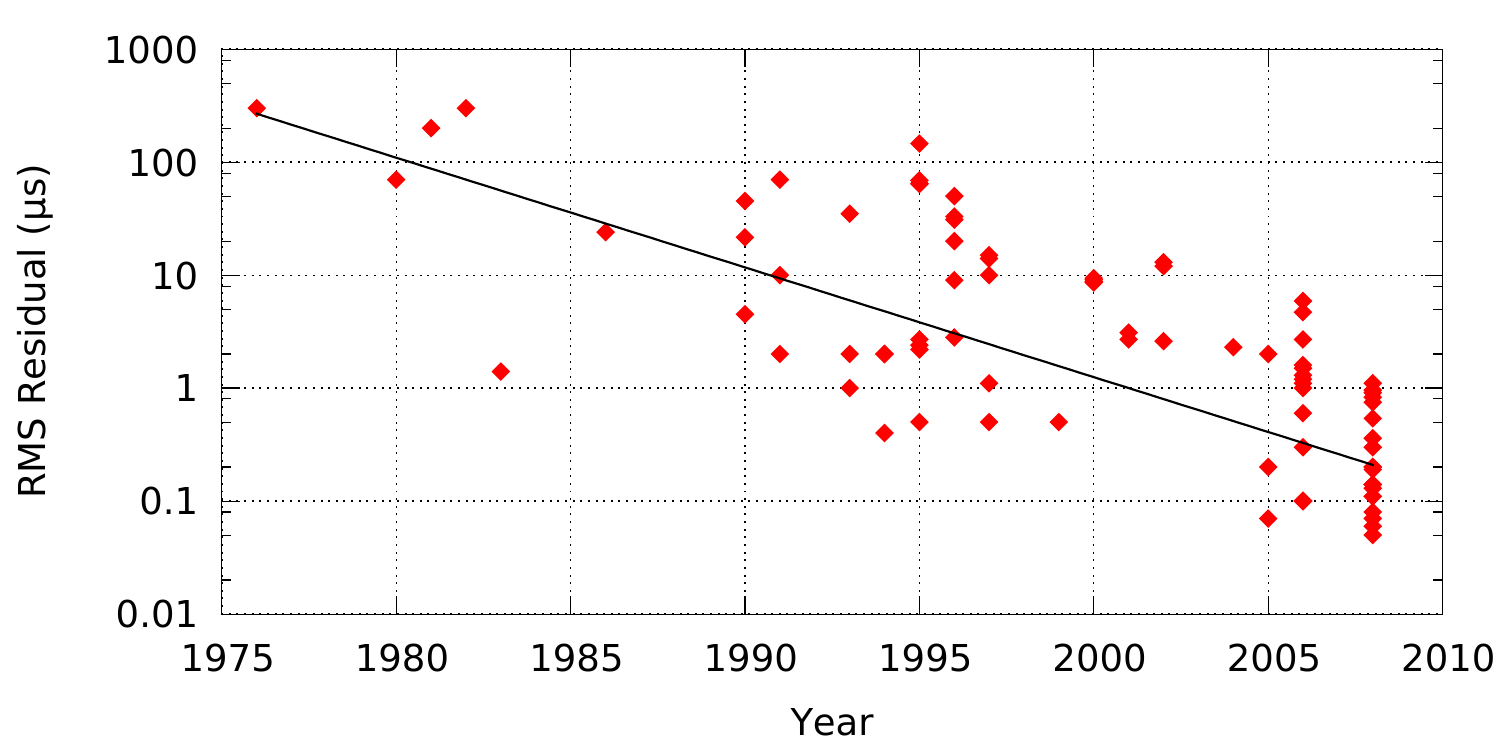}
\end{center}
\vspace*{-6ex}
\caption{\label{fig:rms_vs_time}%
Published RMS pulsar timing residuals versus time, showing exponential
improvement \citep{dj09}.  Continued improvement should allow us to detect GWs
within the next decade.
}
\vspace*{-1ex}
\end{figure}

A single plane gravitational wave with a well defined frequency will 
cause the measured time-of-arrival of individual pulsar pulses to 
oscillate with time; most of our expected signals involve linear superpositions 
of the effects of many such waves.   The NANOGrav goal is to detect such effects 
by careful examination of long-term pulsar timing signatures.  {\bf The single most 
important performance parameter is the root-mean-square (RMS)
``timing residual'' $\sigma_i$ for each of the pulsars in our sample.}  
The fiducial goal of a detection of a stochastic background  
of gravitational waves requires roughly 5 years of timing 20 pulsars, each with
$\sigma_i$ of $\sim$~100~ns \citep{jenet:2005:dsg}.  Our
achievable timing precision has been steadily improving over the last
two decades \citep{dj09}, as shown in
Fig.~\ref{fig:rms_vs_time}.  Given a concerted effort by NANOGrav and its
international collaborators, we expect to continue this trend and achieve
our primary goal --- the detection of GWs in the nanohertz range --- before the end of the decade. 
This will require a substantial amount of observing time on large radio telescopes,
together with computational resources for data analysis and human resources for
algorithm development. 

In this section, the NANOGrav technical program
is outlined in detail, including signal detection, infrastructure requirements and technology drivers.  
References to specific technology driver sections are indicated. 
Current pitfalls are described together
with NANOGrav's plans for mitigating these problems.
The full NANOGrav process can be broken into 4 stages:
observation, time-of-arrival (TOA) calculation, derivation of timing
models and residuals for the individual pulsars, and combining
pulsar residuals to detect or improve upper limits on the GWB.
In addition to the technical issues discussed here, we have another
crucial need:  to recruit and train young scientists at the undergraduate, 
graduate and postdoctoral levels.

\subsection{Observations} \label{sec:obs}

The discovery of millisecond pulsars (MSPs) \citep{backer82} revolutionized
high-precision timing.  MSP rotational phases
 can often be determined within a few $\mu$s or better
in a single observation. The sensitivity of a pulsar timing array (PTA)
is determined by the number and distribution of
the pulsars in the array, the cadence with which they are observed,
and the precision with which the pulse times of arrival are measured.
Currently there are several pulsars for which the RMS timing residual $\sigma_i$ approaches
100~ns  and roughly 20 more with residuals less than 1~$\mu$s.
  
{\bf  Accomplishing our goal of characterizing GW sources
requires routine access to significant amounts of time on large
radio telescopes, plus observing (backend) instrumentation capable of remov-
ing the worst propagation effects and accumulating the pulsar signal over wide
bandwidths. Specifically,  it translates into approximately 10\% of the observ-
ing time on the Arecibo telescope and about 20\% of the time on the Green Bank
Telescope  (see \S\ref{sec:hours}).}

We need large telescopes for this effort since the 
pulsar signal strength increases directly with telescope collecting
area, with a corresponding effect on the GW signal. We also need large 
integration times with large bandwidths since the signal strength is proportional to
the square root
of the product of the observing bandwidth and integration time.
Finally, frequent observing
epochs are necessary: first to ameliorate interstellar propagation effects and
then to prevent long gaps that risk introducing spurious effects into the timing
residuals used for the GW analysis.
Extreme equipment stability and access to the highest quality clock
information are paramount during the observation and between
observation epochs.
Comparison of observational data between observatories
and simultaneous observations with multiple telescopes is also crucial.

Our observation requirements are also dictated by the need to correct
for the propagation of the radio pulsar signal through the interstellar medium
(ISM). This has two major effects on the observed signal.  The first, 
dispersive smearing, can be very well
modeled as a frequency-dependent filter acting on the data stream, and,
hence, be accurately removed by applying the inverse filter
to the pre-detection data.  This coherent dedispersion technique
\citep{hr75}, which is computationally intensive,  
is currently in use across bandwidths of the order of 100~MHz at many of the world's largest telescopes.  
In 2009, it will be
implemented across many hundreds of MHz in the GUPPI
\footnote{Green Bank Ultimate Pulsar Processing Instrument}
at
the GBT, meaning that fully dedispersed pulsar signals will be
available simultaneously across any one of the GBT's sub-3-GHz
receivers.  Similar backends are required for Arecibo (the newly
installed Mock spectrometers have wide bandwidth but cannot perform 
coherent dedispersion) and for future observatories.  

A far more
challenging problem is multipath propagation of the pulsar signal in
the ISM, which causes scintillation of the signal via constructive and
destructive interference at the Earth, and also significant scattering
for many pulsars, especially at low frequencies.  Substantial efforts
are underway to model the multipath effects and find ways to
eliminate them from the pulse profiles (\S\ref{TD:ism}).  Both the
multipath scattering and dispersive effects are observed to change on
timescales as short as one day and as long as decades \citep[e.g.][]{ktr94,cwd+90,Hemberger08}. 
This makes
multifrequency (at least two receivers, each with wide bandwidths, such that
at least 4 separate frequency bands are accessible) observations a requirement at each epoch.

An ever-increasing challenge is the presence of radio-frequency interference (RFI) in the signal. 
Our group has demonstrated initial success with broadband and narrowband software RFI
identification and excision before and during coherent dedispersion \citep{sst+00}, but we need efficient
versions of these algorithms in order to implement them routinely in the wideband
real-time instruments.  Achieving this goal and exploring alternate RFI mitigation schemes
based on FPGA technology \citep{khc+05}
are some of our priorities for the next few years (\S\ref{TD:algorithm}).
We will also determine the exact effect of such techniques on the final recorded 
signal and TOAs by performing an end-to-end simulation of the
pulsar signal path encompassing everything from propagation effects to calibration corrections.

\subsection{Pulse Time of Arrival Determination} \label{sec:toa}

For a given pulsar, the raw data acquired during an observing epoch consists of a set of 
profiles, coherently dedispersed within several channels, detected as self- and cross-polarization 
products, folded synchronously using the best available
ephemeris, and saved to disk roughly every minute.  

Turning these recorded profiles into times of arrival (TOAs) is a fundamental component
of our analysis.  In brief, it involves the cross-correlation of the epoch profile(s) with a standard 
template for the pulsar, and adding the resulting time offset to the recorded clock time.   We have
identified many facets of our current procedures which require significant development.  {\bf In most
cases we know what the starting points for the improvements should be, and we have actually
begun to implement them, but the person-power is currently lacking to accomplish them all
effectively and on the necessary timescale.}  We note that we are developing two 
largely independent software pipelines to facilitate error recognition and correction.
Key items of concern are mentioned below and also highlighted in  \S\ref{TD:algorithm}.

\begin{itemize} 
\item We are implementing 
multiple techniques for computing
the receiver parameters and, and we are comparing these
methods.

\item Standard profile determination is vitally important, as it represents our best guess at the
true envelope of the radio emission from the pulsar, and we are often working to a precision that is $\sim 1$\% of our sampling interval.

\item The TOA determination algorithm most used for the last 15 years has been a frequency-domain
cross-correlation of the total-intensity profile with the standard template \citep{tay92}.  
Although this has worked fairly well, we plan to incorporate full-Stokes information and are considering Bayesian approaches to this problem.
\end{itemize}


\subsection{Timing Model Analysis} \label{sec:timing}

Timing analysis consists of determining the time delay differences (or residuals) $R(t)$ between
the observed TOAs and those predicted by a
multi-parameter timing model.
Parameters going into the timing model always include: information about
clock errors; the
Earth's position and motion in the solar system; the pulsar's rotational frequency
and the first derivative of that frequency; the position, proper motion, and parallax of the
pulsar;  and information about radio frequency delays due to propagation through
the ionized interstellar medium.
In the common case that the MSP is in orbit around another compact
object, full information about the binary orbit -- including post-Keplerian
parameters that, in turn, are a unique probe of general relativity -- is also
needed to specify the timing model.
Development of this timing model for a pulsar is done in a boot-strap fashion over months to
years of observation and analysis.



\subsection{Correlation Analysis}\label{sec:corr}

Because our observing timescales are on the order of years, we will be sensitive to
GWs with frequencies in the nanohertz regime. Such waves would most likely be due to
coalescing supermassive binary black holes. 
 When the timing analysis is applied to
such a signal the dominant term in the residuals would typically be
a cubic polynomial that grows in amplitude and changes in character
as  the time span of the observations increases. 
In the simplest case of a sine wave and a large GW signal from a nearby
binary, it may be possible to detect this signal in the residuals of a single pulsar.
However, the PTA forms a more robust and sensitive detector by allowing searches
for correlations between the residuals from many pulsars distributed across the sky.
The  characteristic gravitational strain to which we are sensitive scales as 
$h_c \propto \sigma/(T^2 \sqrt{N})$, where $\sigma$ is the rms
timing residual, $N$ is the number of pulsars in the array, and $T$ is the
total time span of the observations \citep{jenet:2005:dsg,ktr94}.


In addition to the obvious sensitivity gained from a larger number of pulsars, the additional importance 
of the PTA follows from its ability to distinguish between a gravitational wave signal and the many forms
 of noise Ñ clock errors, solar system ephemeris errors, ISM propagation noise, timing noise intrinsic to pulsars, etc. Ñ that will be present in the data. The unique character of an isotropic background of 
 gravitational waves leads to a well defined expected correlation between the residuals that only 
 depends on the angular separation angle between the pulsars in the array (Hellings \& Downs, 1983; 
 Jenet et al., 2005). Individual gravitational waves will also have a unique correlation signature but with 
 a preferred direction on the sky. By contrast, an error in the reference clock will affect all the pulsars the 
 same way (a monopolar signature), and a position or velocity error in the solar system ephemerides 
 will cause an error that has a preferred direction with a dipolar signature around the sky. Several 
 sources of error, e.g. rotational instabilities of individual pulsars or propagation delays along particular lines of sight, will lead to uncorrelated errors between pulsars.
Due to the necessity of determining the timing model parameters from the 
pulsar data itself, the correlation analysis will be insensitive to GWs at 
particular frequencies (Blandford, Romani, \& Narayan 1984). For example, 
the presence of semiannual (parallax) and annual (positional) terms as 
well as the need to fit for the pulsar period and period derivative 
removes power from the GW signal with periods of 6 months and a year as 
well as removing a linear and quadratic term from the rotational phase as 
a function of time.  In general, frequentist methods have been used to 
search the post-fit timing residuals for the correlated signal expected 
from a stochastic GW background (Jenet et al. 2005) and have set the most 
stringent limits to date on the GW background.  However, accounting for 
the GW power absorbed by the timing model fit has often been problematic 
for these algorithms.  We plan to explore Bayesian approaches to detecting 
and placing limits on the GW background that will more naturally and 
correctly account for these effects (McHugh et al. 1996; Anholm et al. 
2008; van Haasteren et al. 2008).  
Fully developing such methods is a 
cornerstone of our project, since they will lead to more robust upper 
limits, a better understanding of the errors in our data, and ultimately a 
detection of GWs within a shorter time span.

\subsection{New Pulsars for the Timing Array}

To fully characterize the gravitational waveforms and to maximize the scientific
return, more precision MSPs are needed.  Finding pulsars in directions widely
separated from the current set of objects is important since the correlation
over a wide range of separations is the key to detecting
a GW background signal.
While the actual performance of a PTA will depend upon the
specific properties of the pulsars in the \hbox{PTA}, simple test
cases suggest that the significance of a GW detection~$S_{\mathrm{GW}}$
in an $N$-pulsar PTA scales roughly as
$S_{\mathrm{GW}} \propto N$,
provided that all of the pulsars being timed have a sub-microsecond
TOA precision.

Three current searches are
now turning up such objects: the PALFA L-band multibeam survey at
Arecibo, the GBT 350~MHz drift scan survey, and the new Parkes L-band
Digital Survey.  A new GBT low-frequency pulsar
survey would be particularly advantageous because it would increase the
number of Northern hemisphere pulsars.  This area of the sky is
currently under-represented in MSP targets.  Since pulsar flux increases
at low radio frequencies (from $\sim$1~mJy at 1.4~GHz to $\sim$10~mJy at
400~MHz), this survey will identify many new pulsars.  We are also considering a 1.4~GHz GBT survey
using a multibeam receiver.

In all of these searches, advances in computational hardware mean that more complicated signal
processing can be undertaken.  Many millisecond pulsars are in
binaries, a reflection of their origin.  However, finding pulsars in
binaries often requires at least searching for accelerated pulsed
signals, if not a search over binary parameters, in addition to the
standard pulse period and dispersion searches.  Objects in very tight
orbits or a pulsar orbiting a black hole would, of course,
produce spectacular collateral science to the main NANOGrav effort.

\subsection{Justification of Required Telescope Time} \label{sec:hours}
In our Science white paper (Demorest et al 2009) we summarized observing time dedicated to this project
in terms of 100-m telescope equivalent time.  World resources currently provide about 300 100-m
hours per month.
We estimate that at minimum, GW detection requires
$\sim$500 100-m hours per month, based on observing 20 pulsars every 2
weeks, for 3 hours at each of 4 radio frequencies in order to obtain
100-ns or better timing precision.  To begin to characterize the GW sources
requires at least twice as many pulsars \citep{lee:2008:tap} and so 1000 100-m hours of time each month.
Upon purchasing 20\% of the time on the GBT and 10\% of the time at AO, we would have
roughly 1300 100-m hours per month and would be able to characterize the stochastic
GWB as well as continuous and burst sources.  These numbers appear in our cost estimates.

\newpage
\section{Technology Drivers}\label{sec:tech}

\subsection{Algorithm Development}\label{TD:algorithm}

Most of the stages in our technical program (\S3) require improvement of existing
algorithms or development of new ones.  We argue that the entire class of algorithms
developed through the NANOGrav activity (often in concert with our international
collaborators) forms a technological tool kit that will enable science not just with
the world's existing large radio telescopes but also with future telescopes such
as the Square Kilometre Array.  Here we recapitulate the most important algorithms
that require progress.

\begin{itemize}
\item Radio-frequency interference (RFI) excision (\S\ref{sec:obs}), efficiently 
implementing existing software algorithms \citep{sst+00}, and exploring
hardware-based solutions \citep{khc+05}.

\item Characterizing and reducing the effects of interstellar scattering \citep{Hemberger08}
and variable dispersion measures \citep{ktr94,You07} (\S\ref{sec:obs}).  As a relevant
and currently feasible hardware development, we strongly advocate the building of
receivers for the GBT and Arecibo that allow truly simultaneous observations at
two widely spaced frequencies.

\item Pulse profile calibration (\S\ref{sec:toa}) including full instrumental 
characterizations \citep{britton,johnston,van06}.

\item Standard-profile determination (\S\ref{sec:toa}) including frequency 
dependence \citep{lom01}.

\item TOA computational methods (\S\ref{sec:toa}) should include frequency-evolving standard
profiles \citep{lom01} and take advantage of full-Stokes information \citep{van06} and/or Bayesian methods.

\item Pulsar timing algorithms are under continuous international development 
\citep{edwards} (\S\ref{sec:timing}).

\item End-to-end simulation of the pulsar signal data path, including ISM effects, telescope
reception, and our full analysis pipeline (\S\ref{sec:obs})

\item GWB analysis algorithms (\S~\ref{sec:corr}) are still in an early stage of development,
with both frequentist and Bayesian techniques being explored \citep{jenet:2006:ubo,Anholm08,vanHaasteren08}.

\end{itemize}

\clearpage

\subsection{Removing the Effects of the Interstellar Medium}\label{TD:ism}
\begin{figure}[h]    
\begin{center}
\includegraphics[width=\textwidth]{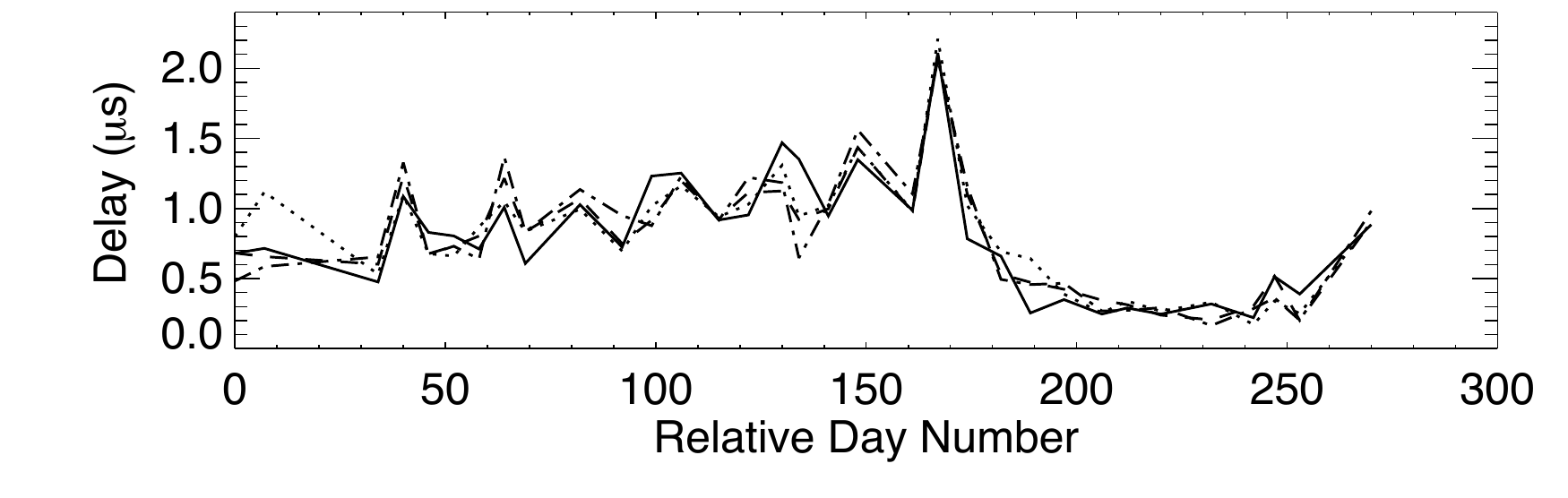}
\end{center}
\vspace*{-6ex}
\caption{\label{fig:hs08}%
Time delay through the ionized interstellar medium toward the
pulsar PSR~B1737+13, estimated using spectral methods at
four separate frequencies near 1400~MHz (adapted from Hemberger and Stinebring 2008).
}
\vspace*{-1ex}
\end{figure}

Although we have made excellent progress in reducing rms residuals
for many sources, we know that time-dependent propagation delays
present a serious challenge in attaining 
100~ns rms residuals toward many pulsars.
We single this out as a technology driver because an extensive
R\&D program will be required, and we are just beginning to
recognize the severity of the problem.
As an example, 
we show in Fig.~\ref{fig:hs08} the results of 270 days of monitoring the
inferred time delay of the signal from one pulsar.
At radio frequencies around 1400~MHz, a commonly used frequency for many
of our high precision timing data, we see the inferred time delay varying from
about $0.2-2\,\mu s$ with several abrupt fluctuations.  



\nocite{Hemberger08}

In an inhomogeneous plasma like the interstellar medium,
 the majority of rays from the pulsar travel along a 
cigar-shaped volume tapered at both ends.
The angular beam waist (scattering radius) is highly wavelength dependent
($\theta_s\propto \lambda^{2.2})$ and varies with time.  Furthermore, this
description of the ray paths (including the exponent of
2.2) depends in a complicated fashion on the
inhomogeneity spectrum and the distribution of material
along the line of sight \citep{cr98}, none of which is known
a priori.

We have considerable expertise on this problem in our group,
and several of our international collaborators are working hard on
the problem, too.
One straightforward approach would be to use spectral information
like that underlying the results in Fig.~\ref{fig:hs08} and simply
subtract the scattering time delay from the TOA.
This requires excellent signal-to-noise ratio, achievable for only
the strongest pulsars and the largest telescopes, and we have
not yet demonstrated that it effectively reduces the rms residuals.
A more ambitious approach, that may be necessary in some cases,
is to use pre-detection information to fully model the phase front of the wave 
as it passes through the interstellar medium.
We have taken some initial steps in this direction (e.g.~Walker and
Stinebring 2005).
\nocite{ws05}
This approach can be thought of as coherent descattering. Unlike
coherent dedispersion, however, this correction requires an
interative modeling approach that, again, requires high sensitivity
and frequent (daily) sampling of the data.
Although the challenges we face in this area are daunting, we note that coherent dedispersion --- one of the current
bedrocks of our observing procedure --- was initially considered farfetched by many, and it has proven to be a highly effective and essential tool for precision pulsar timing.  

\clearpage

\subsection{Pulsar Timing with Interferometers}\label{TD:interferometer}

Over the next decade, a number of new instruments will come on-line,
including the \hbox{EVLA}, the Australian SKA Pathfinder (ASKAP), and
potentially an expanded Allen Telescope Array (ATA-256).  Looking to
the latter half of the next decade and beyond, the Square Kilometre
Array (SKA) is expected to deliver even larger improvements in
sensitivity.  All of these new telescopes are interferometers, in
contrast to the current workhorse telescopes (Arecibo, \hbox{GBT},
and Parkes).

In addition to the possibility of simply additional
sensitivity,\footnote{
The EVLA has an effective aperture equivalent to that of a 130-m
single-dish telescope.}
interferometers offer the possibility of improved scheduling and
efficiency.  In principle, via \emph{sub-arrays}, an array can be
split so that multiple pulsars can be timed simultaneously.
Implementing such a timing program, and achieving an optimum schedule,
will require careful attention to the flux densities of the pulsars in
the PTA and the range of frequencies needed for each pulsar (to
compensate for interstellar scattering effects).

Further, the standard operational mode of interferometers is for
imaging, while pulsar timing represents a \emph{non-imaging}
operational mode.  Rather than visibilities, a \emph{phased-array}
mode must be implemented in (or in addition to) the correlator.  While
conceptually a phased-array mode is not difficult, because pulsar
observing has not been conducted traditionally with interferometers,
the implementation of pulsar-specific processing typically lags that
of standard imaging processing.  As specific examples, implementation
of pulsar modes in the VLBA correlator required approximately 5~years
after the formal dedication of the telescope, and the initial
operational modes of the EVLA correlator will not include pulsar modes.

There is some modest experience with using an interferometer as part
of a \hbox{PTA}, being developed primarily through the EPTA and its
use of the Westerbork Synthesis Radio Telescope (WSRT).  Even this
experience, however, shows that careful attention must be paid to the
process of phasing the array and extracting TOAs from the resulting
phased-array data.

\newpage
\section{Activity Organization}

\paragraph{NANOGrav Internal Organization.} As detailed in \S 3, the
requirements of this project include many aspects ranging from
observing to algorithm development to data analysis, and therefore
necessitate a large collaboration. At present, NANOGrav consists of 25
researchers spread primarily over a host of universities and colleges
across the United States and Canada, serving an ethnically and
socio-economically diverse population of undergraduate and graduate
students. Our specialities include pulsar searching (Cordes, Freire,
Kaspi, Lorimer, McLaughlin, Nice, Ransom, Stairs), timing (Backer,
Demorest, Ferdman, Freire, Gonzalez, Jenet, Kaspi, Lommen, Lorimer,
Nice, Ransom, Stairs, Verbiest), interstellar propagation (Cordes,
Stinebring), gravitational wave detection (Jenet, Lommen),
instrumentation (Demorest, Ransom, Stairs) and polarization (Demorest,
Ferdman, Gonzalez, Rankin). All of these aspects are absolutely
essential to meet our goals. We communicate with each other through
weekly to monthly teleconferences, and collaborative visits. Basic
sharing of data within NANOGrav has been implemented and we are moving
towards common data formats and shared processing algorithms. By
sharing algorithms instead of imposing a single standard, we ensure a
redundancy that allows prompt identification and mitigation of
conceptual and technical errors.

At present, NANOGrav activities are funded through the individual
grants to US and Canadian faculty members as well as various
fellowships awarded to some of the more junior members.  The US
members of NANOGrav recently applied for funds to further strengthen
this collaboration through several measures. One of these is to
increase the number of postdocs and students (both graduates and
undergraduates), since 17 of the 25 current collaborators are senior
researchers. We also plan extensive outreach efforts to involve
high-school students in our research and pave the way for a new
generation of gravitational wave astronomers. We aim to institute a
yearly, week-long winter workshop dedicated to algorithm comparison,
data analysis and resolving of data-related issues. Finally, dedicated
computer hardware will increase the reliability and ease-of-use of our
data sharing infrastructure.

{\noindent\bf International Partnerships:} We organized the first
international pulsar timing array (IPTA) meeting at the Arecibo
observatory in August 2008. This event led to a formal agreement on
collaboration between NANOGrav and its partner organizations in Europe
(the EPTA
) and Australia (the Parkes pulsar timing array or
PPTA
). We currently participate in monthly telecons with those groups and
an agreement on data-sharing has been drafted that will soon enable
regular combination of international timing data. This will allow
rapid identification of technical failures or instabilities at any
telescope and comparison and evaluation of algorithms used at
different institutions. We also plan to involve colleagues in India since
the Giant Meterwave Radio Telescope (GMRT) offers low-frequency
capability that will greatly improve our achievable timing precision.


Along with the PPTA and EPTA, NANOGrav has recently joined the
gravitational wave international committee (GWIC) and has strengthened
its relation with LIGO (two of our current members are also in the
LIGO collaboration). This wider collaboration is needed to fully
benefit from the scientific complementarity of the different GW
detection efforts.

\newpage
\section{Activity Schedule}
\begin{figure*}[b]
  \includegraphics[width=\textwidth]{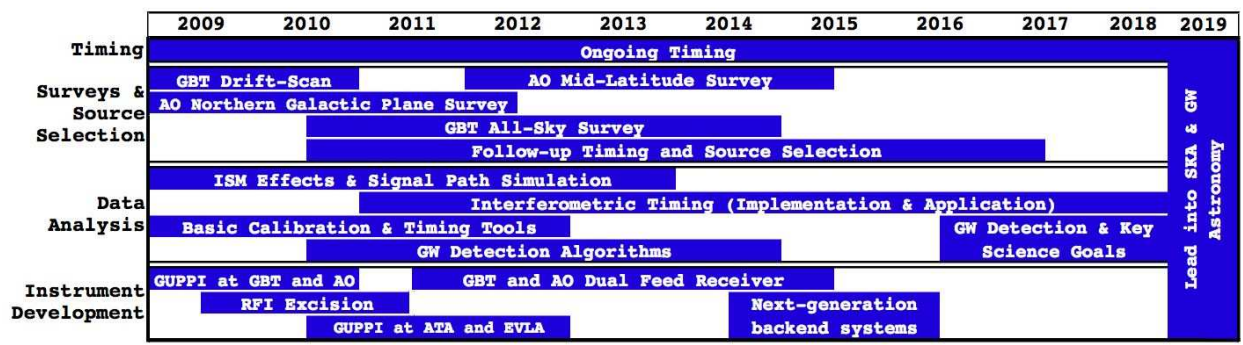}
  \caption{\footnotesize{Timeline for the planned NANOGrav
      activities. A full description is given in the
      text.}}
  \label{fig:Gantt}
\end{figure*} 

\paragraph{Timing Observations} 
As our sensitivity to a gravitational wave background strongly depends
on the length and cadence of the timing data sets, regular and
frequent timing observations will be ongoing throughout the lifetime
of the project, at all available telescopes. 

\paragraph{Source Selection and Pulsar Searches.} 
High precision timing is crucial to achieving our scientific
goals. Since the achievable timing precision varies from pulsar to
pulsar, state-of-the-art surveys are needed to find any rapidly
rotating MSPs to which previous surveys may not have been
sensitive. Planned and currently ongoing searches are:
\def\tightspacing{\topsep=2pt\itemsep=1pt\itemindent=0pt
\labelwidth=11pt\leftmargin=11pt\parsep=0pt}
\begin{list}{$\bullet$}{\tightspacing}
\item\textbf{GBT drift-scan survey:} Observing has finished, processing is
expected to end in 2010.
\item\textbf{Arecibo Northern Galactic plane survey:} Observations are
underway. The survey and data processing are expected to be finished
by
2012.
\item\textbf{GBT all-Northern-sky low-frequency survey:} This survey is
intended to start after the processing of the drift-scan survey is
complete. The survey and analysis should take place from 2010 to
2014.
\item\textbf{Arecibo mid-latitude survey:} This survey would follow the
Northern Galactic plane survey and run from 2012 to 2015.
\end{list}

MSPs detected in any of these surveys may help us to achieve our
scientific goals, but the timing potential of these new pulsars must
be verified. Continuous monitoring will therefore have to be
undertaken for at least a year or more upon discovery. This monitoring
and source selection will hence be ongoing until 2017.

\paragraph{Algorithm Development} 
Four separate development efforts are planned. 
Details on these activities are provided in \S 3 and \S 4.1.
\begin{list}{$\bullet$}{\tightspacing}
\item Basic calibration and timing tools described in \S3.2. We
  anticipate full integration of these algorithms into the current,
  data processing pipeline by the end of 2012, allowing these new techniques to be used in the analysis of the ongoing timing observations from the beginning of the decade onwards.
\item Characterization of the effects of the ISM and the
  implementation of algorithms to mitigate such effects is a
  substantial effort that is expected not to be completed until late
  2013. The second activity in at this stage is the creation of an
  end-to-end simulation of the signal data path will be constructed
  for system-analytic purposes. The construction of this simulation
  and the interpretation of its results are expected to continue into
  2013 as well.
\item GW detection algorithms based on both frequentist and Bayesian
  principles (see also \S 4.1), will be finalized by 2015.
\item As outlined in \S 4.3, interferometers will become increasingly
  important in pulsar timing. Optimization of our techniques and
  algorithms to deal with interferometric telescopes will be
  completed by the latter half of the decade.
\end{list}

\paragraph{Instrument Development}
Over the course of the decade, we expect to make substantial
improvements in pulsar instrumentation, in three phases:
\begin{list}{$\bullet$}{\tightspacing}
\item\textbf{Current backend systems:} The GUPPI backend, currently
under development at Green Bank, will provide high dynamic range, full
polarization, and coherent dedispersion capabilities over an order of
magnitude larger bandwidth than previously possible. By the end of
2009, the full functionality of GUPPI will be implemented. We plan to
install a copy of GUPPI at Arecibo by the end of 2010, and at the ATA
and EVLA by 2012. Concurrently we will investigate the possibility of
adding real-time RFI excision based on established techniques
\citep{sst+00} to these systems.
\item\textbf{Receiver systems:} We will encourage dual-frequency
receivers to be installed at the GBT and Arecibo by 2015. These
receivers will enable simultaneous timing observations at high and low
frequencies. In addition to improving observing efficiency, this will
allow us to better quantify and correct time-variable ISM effects.  We
will also investigate the feasability of a multibeam receiver for the
GBT, which would dramatically improve the speed and sensitivity of
future pulsar surveys.
\item\textbf{Next-generation backend systems:} We plan to continue
research and development of advanced next-generation backend systems,
with a goal of upgrading the current (GUPPI-based) systems by 2015.
The new systems will be designed to make use of the full bandwidth
provided by the new dual-frequency receivers. They will also include
advanced real-time RFI mitigation capability, and the ability to deal
with specialized interferometer configurations such as subarrays or
multiple beams.
\end{list}

\paragraph{Activity Lifetime} 
Contemporaneously with advanced LIGO we expect the PTA sensitivity can reach a level
where we will detect gravitational waves from supermassive black hole
binaries, and detect or place astrophysically significant constraints
on the gravitational signature of early universe phenomena (cf.\
\S\ref{sec:earlyU}). With these observations, we will introduce new
tests of dynamical gravity (cf.\ \S\ref{sec:tests}) and explore black
hole/galaxy co-evolution (cf.\ \S\ref{sec:coevol}). The end of the
2010 decade also matches the projected beginning of the SKA era; our
activities, and those of our international counterparts, will merge
smoothly into intensive timing programs with the SKA and the
commencement of GW astronomy as a science in itself. The efforts
invested now into algorithm development and building up long timing
baselines will ensure that the SKA can produce superb-quality science
from the beginning.

\newpage
\section{Cost Estimates}

In this section, we estimate the cost of running the NANOGrav activity
itself. The operating costs include direct funding to the scientific
community as well as facility costs.

\begin{footnotesize}
\begin{center}
\begin{tabular}{|c|c|c|c|c|}\hline
Item & One-time & Direct Funding & Facilities & Decade total\\ 
     & Direct Funding & Cost/year & Cost/year &             \\ \hline\hline
AO Telescope Time &- & - & 1M & 10M\\\hline
GBT Telescope Time &- & - & 2M & 20M\\\hline
Personnel &- & 2M & - & 20M\\\hline
Computing Hardware & 300K  & 100K & - & 1.3M\\\hline
Travel and Communication &- & 1M & - & 10M\\\hline
Hosting Meetings &- & 300K & - & 3M\\\hline
GBT/AO receiver Upgrade & 1M  &- & - & 1M\\\hline
Recording Hardware Upgrade & 300K &- &- & 300K \\\hline
Next Generation &  & & &  \\
Backend Development & 400K &- &- & 400K \\\hline\hline
\multicolumn{4}{|c|}{Total Cost} & 66M \\\hline
\end{tabular}
\end{center}
\end{footnotesize}

Although we estimate that current levels of time allocated to us from
the Green Bank and Arecibo Telescope will allow a detection of gravitational
waves, more time will be needed to actually characterize gravitational
wave sources.
To achieve the scientific goals we have described in this document we estimate we need
10\% of the observing time at the Arecibo observatory and 20\% at the Green Bank radio
observatory. The value of this is estimated to be about \$2 and \$1
million per year respectively for a total of \$3 million per year. 

Aside from the facilities costs, the major operating cost of NANOGrav will be in supporting
graduate students and a post-doctoral work force.  (We already have the senior staff to supervise
young scientists well in a variety of research and educational settings.)
It is expected that this will
be done through some combination of individual investigator grants
together with a block grant to establish a ``NANOGrav Institute.'' 
We estimate \$2M per year to maintain 8 post-doctoral fellows, 10 graduate
students, 2 hardware engineers, 2 computer scientists and summer salary for 10 faculty.
This includes salary,
overhead and benefits.   
The two Canadian groups also 
contribute personnel via their individual grants and some fellowships to their group members.

In addition to the personnel costs described above, we require funding for computers, travel,
communication, and meetings;  all oriented around organizing and maintaining international
cooperation.

The installation of the GUPPI data taking systems at 
AO, ATA and EVLA will cost about \$100K each for hardware, for a total of \$300K.  Installation costs are included
in the engineering salaries above.
Dual feed receivers for both AO and GBT will cost 
\$500K each for hardware.
Next generation backend development will cost \$100K at each of the 4 telescopes (GBT, AO, ATA, and EVLA) for
a total of \$400K.

Over the entire ten years, the estimated cost to the US of running the
NANOGrav activity will be approximately \$66M (inclusive of the fractional operating costs of the national facilities used by this project). Considering the
cost of other gravitational wave detection efforts, the NANOGrav
activity is remarkably inexpensive.


\bibliography{resp.bib}

\end{document}